\documentclass[aps,prl,preprint,amssymb,groupedaddress]{revtex4}
\usepackage{graphicx}

\begin{document}

\title[]{Reversible Random Sequential Adsorption on a One-Dimensional Lattice}

\author{Jae Woo \surname{Lee}}
\email{jaewlee@inha.ac.kr}
\affiliation{Department of Physics, Inha University Incheon 402-751 Korea}

\begin{abstract}
We consider the reversible random sequential adsorption of
line segments on a one-dimensional lattice. Line segments of length
$l \geq 2$ adsorb on the lattice with a adsorption rate $K_a$,
and leave with a desorption rate $K_d$. We calculate the
coverage fraction, and steady-state jamming limits by a Monte
Carlo method. We observe that coverage fraction and jamming limits
do not follow mean-field results at the large $K=K_a/K_d >>1$.
Jamming limits decrease when the length of the line segment
$l$ increases. However, jamming limits increase monotonically
when the parameter $K$ increases. The distribution of 
two consecutive empty sites is not equivalent to the
square of the
distribution of isolated empty sites.
\end{abstract}

\keywords{random sequential adsorption, jamming limit}
\maketitle

\section{Introduction}
\label{intro}

The irreversible adsorption of large molecules reported on  
systems of colloids, proteins, latex spheres, polymer, 
etc.\cite{PR94,PR97,EV93,FL39,GH74,FG80,OL86,RA93,Lee96,Lee97,Lee03,PF91,WA94}.
The most simple model of the irreversible adsorption is
the random sequential adsorption(RSA). Large molecules impact
sequentially on the surface. If the impacting surface is empty,
the molecules adsorb on the surface and do not detach 
from the surface. If the impacting surface is occupied by molecules,
impacting molecules can not adsorb on the surface. Therefore, we expect
a formation of a monolayer. In the RSA model the coverage fraction of
the surface approaches a limiting value, so called, jamming
limit at the long time. In the lattice model of RSA the coverage fraction
$\theta(t)$ follows a exponential behavior as $\theta(t) =
\theta(\infty) - A \exp(-B t)$ where $\theta(\infty)$ is 
a jamming limit, $A$ and $B$ are constants depending
on the dimensionality of the surface and the shape of  
molecules. In the continuous model of RSA or the
parking-lot problem, the coverage fraction follows a power-law behavior
as $\theta(t) = \theta(\infty) - At^{-\alpha}$
where $A$ is a constant and the exponent $\alpha$ 
depends on the dimensionality and the shape of the object.
The RSA is a oversimplified model for the adsorption of
large molecules. Indeed, there are a lot of effects
such as the transport of molecules, the diffusion of
adsorbed molecules, and the desorption from the surface
to the bulk solution\cite{WA94,SV98,JT94}.

There are many studies on effects of
the desorption on 
the RSA for physical, chemical, and biological systems\cite{PR94,PR97,EV93}.
A simple example of RSA with desorption is a parking lot
problem. Identical cars adsorb(or park) on a line(curb)
at the rate $K_a$. A certain number of parked cars
leave a empty space that is too small to fit another
car. In the irreversible model of the parking lot
(i.e. cars park permanently on the parking lot), the density
of cars reaches to the jamming 
limit $\theta(\infty)=0.747 \cdots$\cite{JT94}.
In the reversible model, the cars park on a line at a rate
$K_a$ and leave the line at a rate $K_d$. In the reversible
model the coverage fraction depends on the adsorption and desorption
rate. For large values of $K=K_a / K_d$ the coverage fraction shows
two different time scales\cite{JT94,KN99,GD20,KB94}.
Jin et.al. studied the adsorption-desorption process of rods
on a line. They observed the logarithmic dependence of the 
coverage fraction\cite{JT94}. However, their approaches are based
on the mean-field idea. Kolan et.al. reported the glassy
behavior of the parking-lot model. They observed  two
different time scales of the coverage fraction by the Monte Carlo
method\cite{KN99}.
Ghaskadvi and Dennin reported the reversible
RSA of dimers on a triangular lattice\cite{GD20}.
Krapivsky and Ben-Naim reported  mean-field results
of the coverage fraction for adsorption-desorption 
processes\cite{KB94}.
They obtained the jamming limit $\theta(\infty) \simeq
1-1/ \log(K), ~(K>>1)$ for the continuum model.
The coverage fraction follows a logarithmic dependence such as
$\theta(t) \sim 1-1/ \log (t) $ for the desorption
controlled limit ($K_a = \infty  , K_d =$finite).
In the lattice model the steady-state coverage fraction
was given as $\theta_{eq}=1-(1/K)^{1/ l} / l ~(K>>1)$
where $l$ is the length of adsorbed objects.

In this work we consider the reversible RSA process
on the one-dimensional lattice by using Monte Carlo
simulation. We observed that the coverage fraction shows
three time scales for large values of $l$.
However, the jamming limit at the steady-state do not
follows the mean-field behavior for the large $K$.
In section II we present the Monte Carlo method of the
reversible RSA. We give results and discussions 
in section III and concluding remarks in section IV.

\begin{figure}
\includegraphics[width=7cm]{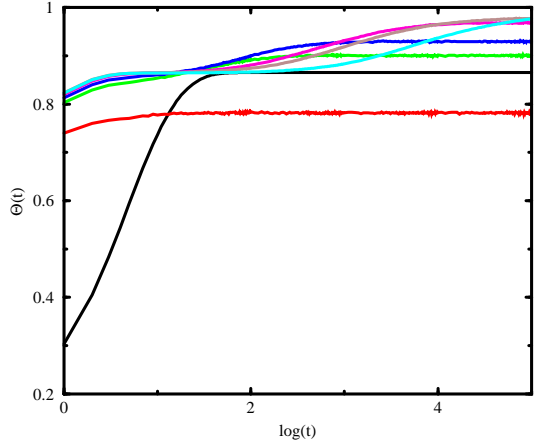}
\caption[0]{
The coverage fraction $\theta(t)$ versus $\log(t)$ for $l=2$
with $K=10, 50, 100, 500, 100, 5000$ from bottom to top, and
$K=\infty$(solid line).
}
\end{figure}

\begin{figure}
\includegraphics[width=7cm]{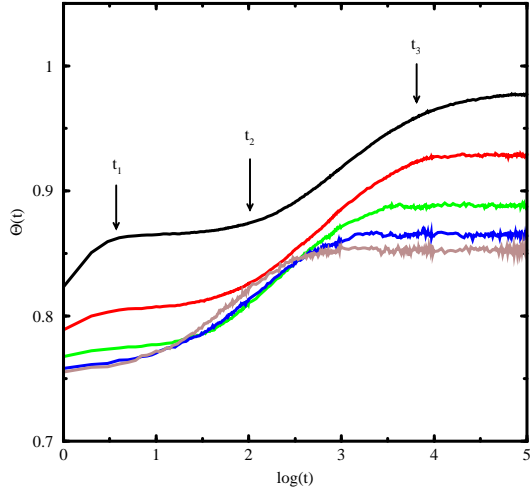}
\caption[0]{
The coverage fraction $\theta(t)$ versus $\log(t)$ for $K=1000$
with $l=2, 4, 8, 16, 32$ from top to bottom.
}
\end{figure}

\begin{figure}
\includegraphics[width=7cm]{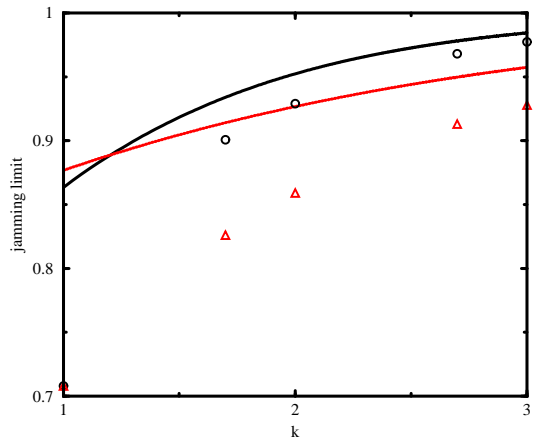}
\caption[0]{
The steady state jamming limit against the parameter $\log(K)$
for $l=2(\circ)$ and $l=4(\triangle)$. Solid lines
represent mean field results.
}
\end{figure}

\section{Reversible RSA model and Monte Carlo Method}
\label{monte}

Consider a one-dimensional clean lattice initially. In the reversible RSA
model, a line segment of a length $l$ drops on the lattice at the
rate $K_a$. 
Adsorbed $l$-mers are desorbed from the lattice at the rate $K_d$. 
We select randomly a lattice site. 
If the selected site is empty, we try to adsorb a line
segment of length $l$ with a probability $p_a = K_a / (K_a + K_d )$.
We generate a random number $p (0 < p \leq 1) $. 
If $p < p_a $, we try to drop a line segment on the lattice.
In the trial of adsorption we check  $(l-1)$ consecutive neighbors.
If  $l$-consecutive lattice sites are all empty,
$l$-mers adsorb on the lattice. Otherwise, the trial is
rejected and increase a Monte Carlo step.
If the selected site is occupied by the adsorbed
line segment, we generate a random number $p$.
If $p>p_a$, the occupied line segment desorbs
from the lattice. Otherwise, the trial of desorption is 
rejected and increase a Monte Carlo step.

It is not allowed to adsorb a line segment on occupied
sites. So we expect a formation of the monolayer.
Kinetics of the coverage fraction
is controlled by a parameter $K= K_a / K_d$. 
We set $K_d =1$ and control the adsorption rate $K_a$.
So that, a line segment will be desorbed from the
lattice every $(K_a + K_d )$-trials. 

We consider a lattice of size $L=10^4$ with a periodic boundary condition.
The lattice is a one-dimensional ring with a length $L$. 
We also check the finite size effect for the large
lattice with $L=10^5$.
We calculate the coverage fraction as 
a function of the time. The coverage fraction was
averaged over 100 different configurations. 
One Monte Carlo time corresponds to $L$ adsorption/desorption trials
regardless of successful or unsuccessful trials.
Maximum Monte Carlo times were up to $t=10^5$. 

\section{Results and Discussions}
\label{resul}

Consider the reversible RSA process of the line segment with the length $l=2$.
In Fig. 1 we present the coverage fraction versus the time. For the irreversible RSA
($K=\infty$), the coverage fraction saturates exponentially to a jamming limit
$\theta(\infty) = 0.868 \cdots$. For a small $K$ 
(for example $K=10$) the jamming limit at the steady-state is smaller
than that of the irreversible RSA. When the parameter $K$ increases
we observed  two different 
linear regions of the coverage fraction when we plot
the coverage fraction $\theta(t)$ against $\log(t)$.
The coverage fraction increases rapidly at the early times $t < t_1$, 
and converges to the jamming coverage fraction
of the irreversible RSA at $t_1 < t < t_2$ as shown in Fig. 2. 
In the former region the adsorption controls the kinetics of the 
RSA. The time $t_1$ and $t_2$ increase when the length of the line
segment $l$ increases as shown in Fig. 2.
At $t_2 < t < t_3$ the coverage fraction increases linearly up to the 
saturation time $t_3$ when we plot the coverage fraction $\theta(t)$
against $\log(t)$. In this region the coverage fraction is proportional
to $\theta(t) \sim \log(t)$.
At $t>t_3$ the coverage fraction saturates to the steady-state jamming limit.
Steady-state jamming values increase when the parameter $K$ increases.

In Fig. 2 we gave the coverage fraction $\theta(t)$ versus $\log(t)$
for a fixed $K=1,000$ and the different length of the line segment $l$.
When the length of the line segment increases, we
observe clear four stage behaviors of the coverage fraction. 
For small values of the line segment $l < 8$, the time $t_1$ is very
small and the coverage fraction approaches to the irreversible jamming value
rapidly. The times $t_2$ and $t_3$ increase as the length of the
line segment increase. Steady-state jamming values 
decreases when the length of the line segment $l$ increases.

In Fig. 3 we present the stead-state jamming limit versus the parameter
$K$ for $l=2(\circ)$ and $l=4 (\square)$. The solid line is 
the mean-field prediction 
$\theta(\infty) = 1- (1/K)^{1/l} / l$ for $K >>1$\cite{KB94}.
Steady-state jamming limits are not coincident with 
mean-field predictions. Monte Carlo results are always smaller
than those of  mean-field results.  Steady-state jamming limits
increases monotonically when the parameter $K$ increases. 

In Fig. 4 we show  steady-state jamming limits against the length
$l$ of line segments.  Jamming limits decrease monotonically when
the length $l$ increases. Mean-field results of  jamming
limits (solid lines in Fig. 4) have a minimum point. However, our
Monte Carlo results do not have a minimum point of the jamming limit. 
The non-mean field behavior of steady-state jamming limits can
be understood by the empty site distribution. 

\begin{figure}
\includegraphics[width=7cm]{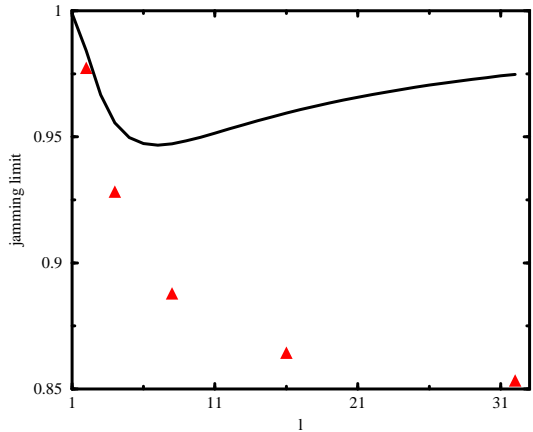}
\caption[0]{
The steady state jamming limit against the length of the line
segment $l$ for $K=1000$. The solid line represent the mean field result.
}
\end{figure}

\begin{figure}
\includegraphics[width=7cm]{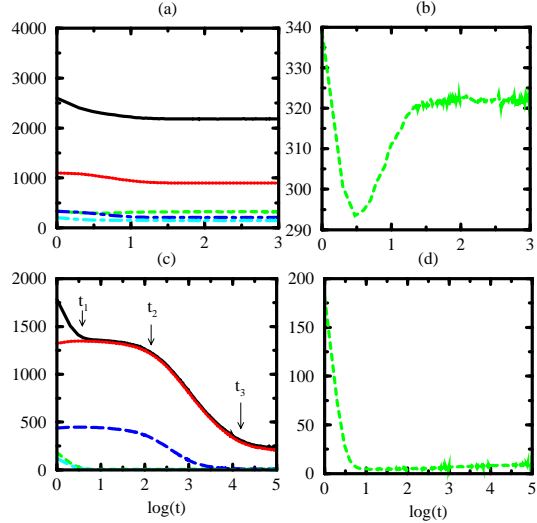}
\caption[0]{
The time dependence of empty site distributions with $l=2$.
We present different kinds of empty site distributions,
the total number of empty sites $N(t)$(solid line), the
number of an isolated empty site $N(x0x,t)$(dotted line), the number
of consecutive two isolated empty sites $N(x00x,t)$(dashed line),
the number of two isolated empty sites separated 
by $l=2$ occupied sites $N(x0xx0x,t)$(longdashed line),
and the number of an isolated empty site and two consecutive
isolated sites separated by $l=2$ occupied sites $N(x0xx00x,t)$(dotdashed line)
(a) for $K=10$, (c) for $K=1000$. Short time behaviors
of $N(x0xx00x,t)$ (b) for $K=10$ and (d) for $K=1000$.
}
\end{figure}

In Fig. 5 we represent
the number of empty sites; 
the total number of empty sites $N(t)$, the number of isolated
single empty sites $N(x0x,t)$, 
the number of two consecutive empty sites $N(x00x,t)$,
the number of isolated empty sites separated by $l=2$ occupied sites
$N(x0xx0x,t)$, and the number of an isolated empty site and
two consecutive empty sites separated by $l=2$ occupied sites
$N(x0xx00x,t)$ where $x$ means an occupied site and $0$ denotes an
empty site. The total number of empty sites
$N(t) =(1-\theta(t))L $ saturates to  steady-state limits at 
long times as shown in Fig. 5 (a) and (c).
The early time behavior
of the empty site is controlled by  collective behaviors of the
single empty site,  double empty sites, and higher empty sites. 
The number of two consecutive empty sites $N(x00x,t)$ shows a 
minimum point around the time $t_1$ and saturates to a steady-state value
at $t> t_1$ as shown in Fig. 5 (b) and (d). In particular, we observe that
$N(x0xx0x,t) \neq N(x0x,t)N(x0x,t)$ and 
$N(x0xx00x,t) \neq N(x0x,t)N(x00x,t)$. The mean-field prediction
is based on the approximation $N(x0xx0x,t)=N(x0x,t)^2$ and
$N(x0xx00x,t) = N(x0x,t)N(x00x,t)$. We conclude that
the coverage fraction of the reversible RSA can not explain by the
mean-field approximation because there is strong
collective behaviors of  empty site distributions.  For large 
$K$(for example $K=1000$)  the total number of empty sites $N(t)$ are controlled by
$N(x0xx0x,t)$ at the early time $t < t_1$.

In Fig. 5 (c) we also observe  typical four stages of the coverage fraction.
At $t<t_1$ the coverage fraction increases due to the dominant process of 
the adsorption. At $t_1 < t < t_2$ the coverage fraction saturates to
the jamming limit of the irreversible RSA. In these region 
$N(x0x,t)$ and $N(x0xx0x,t)$ decays slowly up to time $t_2$ as shown
in Fig. 5 (c). At $t_2 < t < t_3$ the coverage fraction increases
as $\theta(t) \sim \log(t)$ and $N(t) \sim (\log t)^{-1}$. 
In this region isolated single empty sites $N(x0x,t)$
and $N(x0xx0x,t)$ decrease logarithmically, while
$N(x00x,t)$ decreases very slowly and saturates to a steady-state
value. At $t > t_3$  empty sites distribution converges to a 
limiting value
and the empty sites strongly depend on  isolated empty sites
at $t>t_3$.

\section{Conclusions}
We have observed that the coverage fraction of the reversible
RSA does not show the mean-field behavior on the one-dimensional 
lattice. Jamming limits of the coverage fraction decrease
monotonically for the length of the line segment $l$.
The distribution of  two consecutive 
empty sites is not production of the distribution
of  isolated single empty sites.

\begin{acknowledgements}
This work has been supported by Inha University Grant(Inha-30205).
\end{acknowledgements}


\newcommand{\jpa}{J. Phys. A}

\end{document}